\documentclass[onecolumn,amsmath,amssymb,12pt,superscriptaddress,nofootinbib]{revtex4}
\pdfoutput=1

\usepackage[latin1]{inputenc}
\usepackage[english]{babel}
\usepackage{amssymb}
\usepackage{amsmath}
\usepackage{amsthm}
\usepackage[]{graphicx}
\usepackage[]{subfigure}
\usepackage{tensor}
\usepackage{color}
\usepackage{cancel}
\usepackage{setspace}
\usepackage{fancyhdr}
\usepackage[bookmarks,linktocpage, colorlinks=true, plainpages = false, citecolor = blue,  linkcolor=blue, urlcolor = blue, filecolor = blue]{hyperref} 

\begin{document}

\allowdisplaybreaks
\begin{titlepage}

\title{Allowable Complex Scalars from \\ Kaluza-Klein Compactifications and Metric Rescalings \vspace{.1in}}

\author{Jean-Luc Lehners \vspace{.1in} \\ {\it Max--Planck--Institute for Gravitational Physics (Albert--Einstein--Institute) \\ 14476 Potsdam, Germany \\ jlehners@aei.mpg.de}}


\begin{abstract}
\vspace{.2in} \noindent 

Recently there have been discussions about which complex metrics should be allowable in quantum gravity. These discussions assumed that the matter fields were real valued. We make the observation that for compactified solutions it makes sense to demand convergence of the theory's path integral in the higher-dimensional parent theory. Upon compactification this allows for more general matter configurations in the lower-dimensional theory, in particular it allows for complex scalar fields, with a bound on their imaginary parts. Similar considerations apply to metric rescalings in the presence of higher curvature corrections. We illustrate this effect with the example of the no--boundary proposal, in which scalar fields are typically required to take complex values. We find that complex no-boundary solutions exist, and satisfy the derived bound, if the potential is sufficiently flat. For instance, for a compactification from $D$ dimensions, the bound on the imaginary part $\textrm{Im}\,(\phi)$ of the internal volume modulus reads $V_{,\phi}/V < \sqrt{\frac{D-4}{D-2}}/3\sqrt{2}.$ This leads to a mild tension with swampland conjectures.

\end{abstract}
\maketitle

\end{titlepage}

\tableofcontents


\section{Introduction}

Euclidean and, more generally, complex metrics have proven useful in a number of contexts: for example, Wick rotated versions of black hole metrics typically provide the quickest way to derive thermodynamic properties of black holes \cite{Gibbons:1976ue}. Moreover, quantum effects in gravity are often described by analytically continued metrics, such as wormholes \cite{Hebecker:2018ofv}, Coleman-DeLuccia instantons \cite{Coleman:1980aw}, or the bubble of nothing \cite{Witten:1981gj}. However, it is also clear that not all complex metrics make sense. Counter-examples include zero-action wormholes \cite{Witten:2021nzp}, or unstable quantum bounces \cite{Bramberger:2017cgf}. Thus, one needs a criterion specifying which complex metrics should be allowed, and which not.

This question has gained renewed interest recently, due to a proposal of Kontsevich and Segal \cite{Kontsevich:2021dmb}, building on earlier work of Louko and Sorkin \cite{Louko:1995jw}, in the context of quantum field theories (without dynamical gravity). Their proposal is to allow for all complex metrics on which the path integral over all possible matter fields converges. Here the matter fields are taken to be scalars and $p$-forms of all possible ranks, and moreover these matter fields are assumed to take real values only. One reason for this assumption is that it allows for a context-independent definition of matter states. Explicitly, the condition reads
\begin{align}
& |e^{\frac{i}{\hbar}S}|<1 \,\,\, \textrm{or} \,\,\, |e^{-\frac{1}{\hbar}S_E}|<1 \,\,\, \textrm{implying} \\  & Re\left[ \sqrt{g} g^{j_1 k_1} \cdots g^{j_{p+1} k_{p+1}} F_{j_1 \cdots j_{p+1}} F_{k_1 \cdots k_{p+1}}\right] > 0\,, \label{KS}
\end{align}
where $S_E$ is the Euclidean action and \eqref{KS} gives the implied condition stemming from the kinetic term of a $(p+1)-$form field strength $F$. If one now writes the metric in diagonal form (which can always be done locally),
\begin{align}
g_{jk} = \delta_{jk} \lambda_j
\end{align}
then one may show that convergence (for all $p$) imposes a bound on the sum of arguments of the diagonal metric elements \cite{Kontsevich:2021dmb}
\begin{align}
\Sigma \equiv \sum_{j=1}^D |Arg(\lambda_{j})| < \pi\,. \label{bound}
\end{align} 
Note that Euclidean metrics satisfy the bound trivially, with $\Sigma=0,$ while Lorentzian metrics saturate the bound, with $\Sigma=\pi.$ This is consistent with the fact that real-time Feynman path integrals are only conditionally convergent, see e.g. \cite{Feldbrugge:2017kzv}. Thus, as emphasised in \cite{Kontsevich:2021dmb}, Lorentzian physics takes place on the boundary of the domain of allowable metrics.

Witten applied the criterion \eqref{bound} to theories with dynamical gravity \cite{Witten:2021nzp}, noting that it indeed eliminates pathological examples, and retains the above-mentioned ``good'' examples such as analytically continued black hole spacetimes. In \cite{Lehners:2021mah,Jonas:2022uqb} this program was extended to cosmology and to considerations of integration contours for the path integrals. These works showed that no-boundary metrics (without scalar fields) pass the allowability criterion in all dimensions, while proposed examples of quantum bounces do not.

Thus the Louko-Sorkin-Kontsevich-Segal-Witten (LSKSW) bound is empirically found to be rather successful, though one may wonder whether it is too strong in the gravitational context. One indication comes from cosmology: when discussing early universe models, one notices that gauge invariant curvature perturbations mix metric and scalar field degrees of freedom \cite{Bardeen:1980kt}. This makes one question whether it makes sense to treat the two types of fields differently. Another indication is that theories of gravity can include a scalar degree of freedom, in particular $f(R)$ theories \cite{Stelle:1977ry}. In that context the scalar is not fundamental, rather it arises from a rescaling of the metric. But if the metric can take on complex values, then there appears to be no need to require this scalar to take real values. Perhaps the clearest argument is to be found in the context of supergravity theories. In theories with extended supersymmetry, the multiplets containing the graviton also contain vectors and, with enough supersymmetry, scalars. Moreover, supergravity theories are linked to each other by a web of compactifications \cite{Duff:1986hr}. Dimensional reduction gives rise to many vectors and scalars that were originally part of the metric in the higher-dimensional theory. Meanwhile, higher-dimensional form fields give rise in the lower dimensions both to form fields of the same rank and to those with progressively smaller rank, as indices can take values in the internal dimensions. All of this suggests that the allowability criterion should really be imposed in the higher-dimensional parent theory. 

The maximal supergravity theory is found in $11$ dimensions \cite{Cremmer:1978km}. As bosonic fields, it contains gravity and a $3-$form gauge potential with $4-$form field strength. It would thus make sense to impose convergence of the path integral weighted by the corresponding action, with the $4-$form taking real values. Upon dimensional reduction, convergence criteria for the lower-dimensional theories follow. Once one gets down to $8$ dimensions, one obtains form fields of all ranks (for example, the $3-$form gauge potential gives rise to a scalar when all its indices are pointing in the internal dimensions). Thus from $8$ dimensions downwards, one formally recovers the same bound \eqref{bound}. 

Our main observation is that nevertheless, upon compactification, this does not translate into the original LSKSW criterion in $4$ dimensions, because of the fact that some lower-dimensional fields are actually part of the higher-dimensional metric, and are therefore not required to take on real values. We will illustrate this both with examples of compactification (section \ref{sec:compact}) and of rescaling of the metric (section \ref{sec:rescaling}). No-boundary solutions provide an ideal test bed for these ideas, as they typically require scalar fields to take complex values in order to keep the metric and matter configurations regular (section \ref{sec:nobdy}). As we will discuss, convergence of the parent theory then puts a bound on the imaginary part of the induced scalar fields, and this bound is satisfied as long as the scalar potential is flat enough. The bound that is found, and which we will derive in detail for specific examples, roughly speaking requires the potential to be flat enough to allow for inflationary solutions to exist. In the discussion section \ref{sec:discussion}, we will comment on possible implications of this result.

\section{Kaluza-Klein volume reduction} \label{sec:compact}

We consider a compactification from $D$ down to $d$ dimensions, with $d>2,$ on an internal $(D-d)-$dimensional manifold with volume determined by a scalar field $\phi.$ We would like to specify a compactification ansatz such that $\phi$ is canonically normalised in $d$ dimensions. We will focus on the dimensional reduction of the Einstein-Hilbert action. Capital Latin indices run over $D$ dimensions, small case Latin indices over the internal manifold, and Greek indices over the $d$-dimensional external manifold. (A detailed description of Kaluza-Klein compactifications can be found e.g. in \cite{Duff:1986hr}.)

Write the compactification ansatz as 
\begin{align}
g_{MN} dx^M dx^N = e^{2a\phi} g_{\mu\nu}dx^\mu dx^\nu + e^{2b\phi} g_{ij} dx^i dx^j\,, \label{KKansatz}
\end{align}
where $a,b$ remain to be determined. Here we are assuming the coordinate dependencies $g_{\mu\nu}(x^\rho), g_{ij}(x^k)$ and $\phi(x^\mu).$ Indicating higher-dimensional quantities by hats, one then finds that 
\begin{align}
\sqrt{-\hat{g}}\hat{R} = e^{[(d-2)a +(D-d)b]\phi} \sqrt{-g}R + \cdots
\end{align}
If we would like to end up in Einstein frame in $d$ dimensions, then we must choose
\begin{align}
a=-\frac{D-d}{d-2}b\,.
\end{align}
With this choice, one obtains
\begin{align}
\hat{R}_{\mu\nu} & = R_{\mu\nu} + \frac{D-d}{d-2} b\, g_{\mu\nu} \, \Box\phi - b^2 \frac{(D-d)(D-2)}{d-2} \phi_{,\mu} \phi_{,\nu}\,, \\
\hat{R}_{\mu i} & = 0 \,, \\
\hat{R}_{ij} & = R_{ij} - b \, e^{2(b-a)\phi} \, g_{ij}\, \Box\phi \,.
\end{align}
In order for $\phi$ to be canonically normalised, we must therefore set
\begin{align}
b=\pm \sqrt{\frac{d-2}{(D-d)(D-2)}}\,,
\end{align}
and we will choose the plus sign above (this sign is not physically significant). At the level of the action, the dimensional reduction thus yields
\begin{align}
    \frac{1}{2}\int d^Dx\sqrt{-\hat{g}}\hat{R}=\frac{Vol}{2}&\int d^dx\sqrt{-g} \left[R- \nabla_\mu\phi\nabla^\mu\phi+e^{-2\sqrt{\frac{D-2}{(D-d)(d-2)}}\phi}\bar{R}\right]\,,
\end{align}
where $Vol$ is the volume of the unit-radius internal space and $\bar{R}$ the integral of its curvature scalar. The last term is a potential for $\phi$ (further contributions typically arise from the dimensional reduction of other terms in the $D-$dimensional theory). In order for this term to make sense in $d$ dimensions, one usually takes the internal metric $g_{ij}$ to be real valued. If that is the case, then we can see from \eqref{KKansatz} that we get an extra contribution to the LSKSW sum $\Sigma$ of $(D-d) 2\sqrt{\frac{d-2}{(D-d)(D-2)}}|\textrm{Im}\,(\phi)|.$ Note that it is the imaginary part, and not the argument, of the scalar $\phi$ that contributes to $\Sigma.$ This comes from the normalisation of the scalar field.

In order to see how the allowability criterion is affected, one has to make assumptions about the metric $g_{\mu\nu}.$ If the lower-dimensional metric is Euclidean, then we get a total contribution of 
\begin{align}
\Sigma = \left[d\sqrt{\frac{D-d}{(D-2)(d-2)}} + \sqrt{\frac{(D-d)(d-2)}{D-2}}\right] 2 |\textrm{Im}\,(\phi)|\,.
\end{align}
For cases in which the higher-dimensional theory must satisfy the LSKSW bound $\Sigma < \pi$, we thus find a restriction on how large the imaginary part of $\phi$ can be. For instance, if we compactify from $D$ down to $4$ dimensions, we find
\begin{align}
6\sqrt{2}\sqrt{\frac{D-4}{D-2}}|\textrm{Im}\,(\phi)| < \pi\,, \quad \textrm{or} \quad |\textrm{Im}\,(\phi)| < \sqrt{\frac{D-2}{D-4}}\frac{\pi}{6\sqrt{2}} \approx \frac{\pi}{10}\,. \label{Euclibound}
\end{align}
We can see that the imaginary part of $\phi$ is required to be quite small, but crucially it may be non-zero. This extends the LSKSW criterion in the lower-dimensional theory.

If the lower-dimensional metric $g_{\mu\nu}$ is Lorentzian, then the contributions from the time-time and a spatial-spatial components will cancel, leaving
\begin{align}
\Sigma = \pi + 4\sqrt{\frac{(D-d)(d-2)}{D-2}} |\textrm{Im}\,(\phi)|\,.
\end{align}
Then $\phi$ must be real already in order for the metric to sit at the boundary of the allowed domain.


\section{Conformal transformations} \label{sec:rescaling}

Another situation of interest arises when quantum corrections to the Einstein-Hilbert action are present. These arise generically, and take the form of terms that involve higher powers of the Riemann curvature tensor. In quantising (super)gravity theories, these terms arise as counter terms. In string theory, they arise due to the finite length of the string, as $\alpha^\prime$ corrections. 

Let us then consider the example of a gravity theory augmented by higher curvature corrections, for simplicity involving only the Ricci scalar $\hat{R},$
\begin{align}
    S &= \frac{1}{2}\int d^{D} x \sqrt{-\hat{g}} f(\hat{R}) \,,
\end{align}
where $f$ is a function that we may think of as a Taylor series, $f=\sum_{m \in \mathbb{N}} f_m \hat{R}^m.$ The terms $\hat{R}^m$ with $m \geq 2$  effectively introduce a new scalar degree of freedom \cite{DeFelice:2010aj}. This is most easily seen by performing a conformal transformation on the metric, 
\begin{align}
    \hat{g}_{\mu\nu} \equiv e^{\frac{2}{\sqrt{(D-1)(D-2)}}\phi} g_{\mu\nu}\,, \label{metricconformal}
\end{align}
under which 
\begin{align}
    \sqrt{-\hat{g}} = e^{\frac{D}{\sqrt{(D-1)(D-2)}}\phi} \sqrt{-g}\,, \qquad \hat{R} = e^{-\frac{2}{\sqrt{(D-1)(D-2)}}\phi}\left[R  - \nabla^\mu \phi \nabla_\mu \phi - 2\sqrt{\frac{D-1}{D-2}}\Box\phi\right]\,.
\end{align}
The specific numerical coefficient in \eqref{metricconformal} is again chosen such that the scalar field is canonically normalised.
The trick now is to rewrite the action as
\begin{align}
    S = \frac{1}{2}\int d^D x \sqrt{-\hat{g}} \left[ f_{,\hat{R}} \hat{R} - U \right] 
\end{align}
with $U = (f_{,\hat{R}}\hat{R}-f).$ Then one finds
\begin{align}
    S = \frac{1}{2}\int d^D x \sqrt{-g} & \left\{  f_{,\hat{R}} e^{\sqrt{\frac{D-2}{D-1}}\phi}\left[ R  - \nabla^\mu \phi \nabla_\mu \phi - 2\sqrt{\frac{D-1}{D-2}}\Box\phi \right]  - e^{\frac{D}{\sqrt{(D-1)(D-2)}}\phi} U \right\} \label{eq1}
\end{align}
In order to end up in Einstein frame we should now link the scalar to the derivative of the action, according to
\begin{align}
e^{\sqrt{\frac{D-2}{D-1}}\phi} = f_{,\hat{R}}\,.    \label{phiR}
\end{align}
Here we see that it is the $\hat{R}^m$ terms with $m \geq 2$ that give rise to the scalar mode, as they lead to a non-trivial derivative $f_{,\hat{R}}$. The identification above has the consequence that the $\Box \phi$ term in \eqref{eq1} turns into a total derivative and can be dropped. The potential is then given by
\begin{align}
    V(\phi) = \frac{1}{2}e^{\frac{D}{\sqrt{(D-1)(D-2)}}\phi} (f_{,\hat{R}}\hat{R}-f)\,,
\end{align}
where the inverse transformation $\hat{R}(\phi)$ is implicit. 

How complex can the scalar be? Again, we must study this case by case, making assumptions about the metric. If the re-scaled metric is Euclidean, then the sum over arguments is
\begin{align}
\Sigma = D |\textrm{Arg}\, e^{\frac{2}{\sqrt{(D-1)(D-2)}}\phi}| = \frac{2D}{\sqrt{(D-1)(D-2)}}|\textrm{Im}\,(\phi)|
\end{align}
If we impose the LSKSW bound and require $\Sigma < \pi,$ then we obtain the bound
\begin{align}
|\textrm{Im}\,(\phi)| < \frac{\sqrt{(D-1)(D-2)}}{2D}\pi \approx \frac{\pi}{3}\,, \label{rsb}
\end{align}
where the (crude) approximation above assumes the standard range of interest, $4 \leq D \leq  11.$ Once again note that it is the imaginary part of the scalar, rather than its argument, that is bounded. This can thus be a significant restriction on solutions. 

We will now look at specific examples to illustrate these considerations. The no-boundary proposal provides an ideal test bed for this.


\section{Complex scalars in the no-boundary proposal} \label{sec:nobdy}

The no-boundary proposal \cite{Hartle:1983ai,Hartle:2008ng} provides a context in which the scalar field is generically required to be complex valued. We will illustrate this by estimating how large the imaginary part of the scalar typically has to be, generalising a calculation of Lyons who considered a potential consisting of a mass term \cite{Lyons:1992ua} (see also \cite{Janssen:2020pii} for closely related work). Here we will extend this analysis to an arbitrary inflationary potential. As we will see, our enlarged allowability criterion nevertheless leads to a constraint on inflationary potentials. For a detailed discussion of how to define the no-boundary wave function in the presence of a scalar field, we refer the reader to \cite{Jonas:2021ucu}.

We will assume that the lower-dimensional theory resides in $d=4$ dimensions. Further, we will assume that it contains a scalar field with a positive potential $V(\phi).$ Such a model can arise for example from compactification of a higher-dimensional model including higher derivative $\alpha^\prime$ corrections, and flux terms \cite{Ketov:2017aau,Otero:2017thw}. Assuming spatial isotropy and homogeneity, the equations of motion and constraint are given by
\begin{align}
0 &= \phi'' + 3 \frac{a'}{a}\phi' - V_{,\phi}\,, \\
0 &= 3  a'' + a(V+\phi'{}^2) \,, \\
0 &= 3  a'{}^2 - 3 + a^2 (V-\frac{1}{2}\phi'{}^2)\,,
\end{align}
where the scale factor is denoted $a(\tau),$ with $\tau=it$ Euclidean time and primes denoting derivatives w.r.t. $\tau.$  The main feature of a no-boundary solution is that it remains regular as the scale factor vanishes, a locus often referred to as the 'South Pole' of the solution. Shifting the origin of time such that  $a(\tau=0)=0,$ we can expand the equations of motion around the origin to find
\begin{align}
a(\tau) &= \tau -\frac{V}{18}\tau^3 + \frac{8V^2-27V_{,\phi}^2}{8640} \tau^5 + {\cal{O}}(\tau^7)\,, \\
\phi(\tau) & = \phi_{SP} + \frac{V_{,\phi}}{8}\tau^2 + \frac{2VV_{,\phi}+3V_{,\phi}V_{,\phi\phi}}{576} \tau^4 + {\cal{O}}(\tau^6) \,, \label{expansionphi}
\end{align}
where all quantities related to the potential are evaluated at the South Pole value $\phi_{SP}.$ Our goal will be to estimate $\phi_{SP},$ which is the only parameter in the solution.

\begin{figure}[h]
	\centering
	\includegraphics[width=0.8\textwidth]{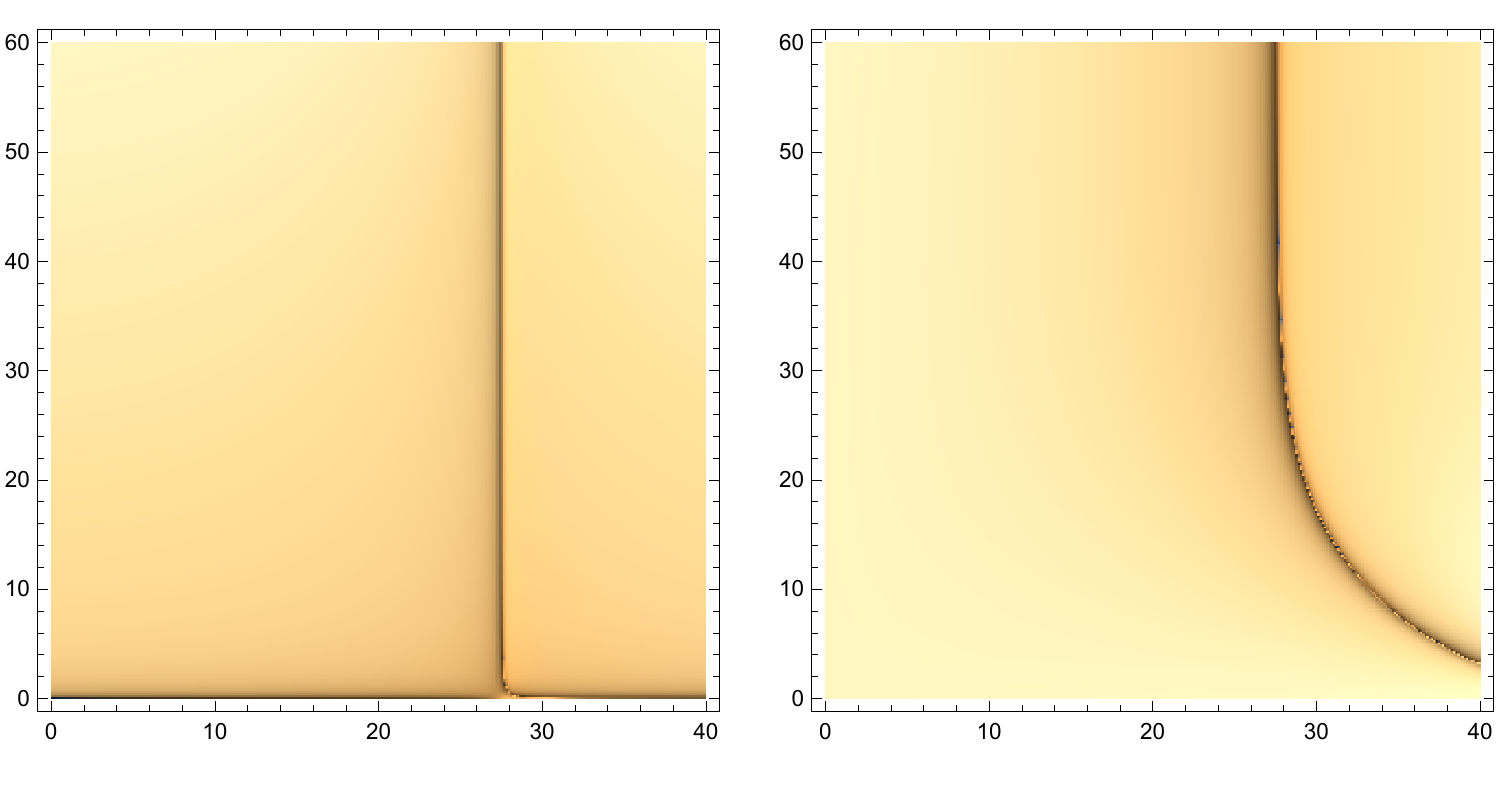}
	\caption{Imaginary field values for $a$ (left) and $\phi$ (right) respectively plotted over a region of the complexified time plane (Euclidean time is in the horizontal direction, Lorentzian time vertical). Dark lines correspond to zero imaginary part, i.e. to the locus of real field values.For this example, the potential was chosen to be of the form $V(\phi)=\frac{1}{100}-e^{-\phi}.$ The initial condition is $\phi_{SP}=8.0700 -0.050189i,$ and the solution reaches the real values $a_1=170,\phi_1=8$ at 'time' $\tau=27.642+52.093i.$}
	\label{fig:excomplextime}
\end{figure}

\begin{figure}[h]
	\centering
	\includegraphics[width=0.4\textwidth]{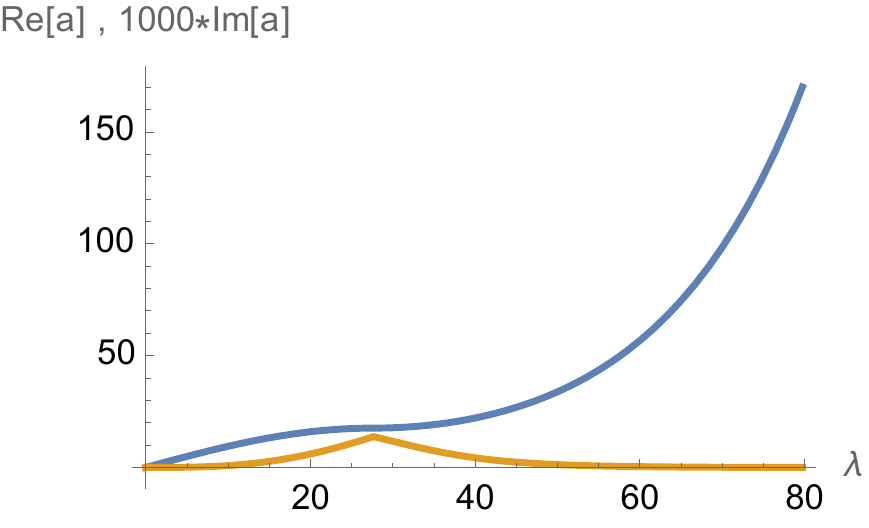}
	\includegraphics[width=0.4\textwidth]{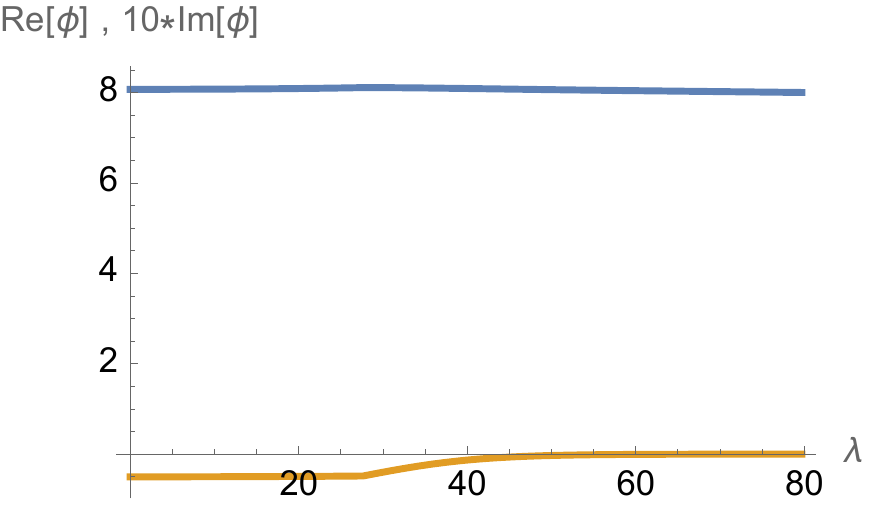}
	\caption{Field evolutions for the example shown in Fig. \ref{fig:excomplextime}, along a contour running from the South Pole in the Euclidean direction first, and then in the Lorentzian direction. Note that the imaginary values (in orange) have been magnified to improve visibility. One can thus see that along this contour the scale factor is real to high precision, while the scalar field is initially complex and then approaches real values along the Lorentzian direction.}
	\label{fig:exfields}
\end{figure}

An inflationary no-boundary solution has recognisable characteristics \cite{Hartle:2008ng,Battarra:2014kga}. Fig. \ref{fig:excomplextime} shows a numerical example of such a solution. The graphs show the imaginary parts of the scale factor $a$ (left panel) and the scalar field $\phi$ (right panel) in the complex time plane, where the dark lines indicate zero imaginary part, i.e. real field values. The horizontal direction denotes Euclidean time, and the vertical direction Lorentzian time. Although the South Pole value $\phi_{SP}$ can be complex, a no-boundary solution is only physical if it reaches real field values $a_1, \phi_1$ at a late time. As the figure indicates, the scale factor is approximately real along a line segment emanating in the Euclidean direction from the origin, and then along a vertical line segment extending to late Lorentzian values. Meanwhile, the scalar field asymptotically approaches real field values along the same Lorentzian time direction. We can use these observations to determine $\phi_{SP}.$ We will assume $|Im(\phi)| \ll 1,$ an assumption that we will see is justified. Note moreover that the metric is Euclidean near the South Pole, and that thus the main contribution to the allowability function $\Sigma$ comes from the scalar field. As the graphs already indicate -- see the right panel in Fig. \ref{fig:exfields} -- the scalar field takes its largest imaginary part at the South Pole, so it is here that we can determine whether a no-boundary solution is allowable or not. Away from the South Pole, suitable complex time paths (suitable in the sense that they keep $\Sigma$ small) can be found using the methods presented in \cite{Jonas:2022uqb}; these paths tend to be close to the Euclidean-followed-by-Lorentzian contour used to produce Fig. \ref{fig:exfields}.

The solution for the metric along the Euclidean axis is rather simple, and corresponds to half of a $4$-sphere,
\begin{align}
a(\tau) \approx \sqrt{\frac{3}{V}}\sin\left( \sqrt{\frac{V}{3}}\tau\right)
\end{align}
where $V=V(\phi_{SP})$ is approximately real. The equator of the sphere is reached at $\tau_{max}^R=\sqrt{\frac{3}{V}}\frac{\pi}{2}.$ Here we denote real and imaginary parts by $R,I$ superscripts respectively. Meanwhile, the scalar field is roughly constant along the Euclidean axis, reaching the value (cf. \eqref{expansionphi})
\begin{align}
\phi(\tau_{max}) \approx \phi_{SP} + \frac{3\pi^2}{32}\frac{V_{,\phi}}{V}\,.
\end{align} 
This equation shows us that the accuracy of our approximations is on the order of ${\cal O}\left(\frac{V_{,\phi}}{V}\right) = {\cal O}(\sqrt{\epsilon}),$ where $\epsilon \equiv \frac{V_{,\phi}^2}{2V^2}$ is the slow-roll parameter. For realistic models, the accuracy is thus at the level of a few percent.

At late times, we can approximate the solution with the slow-roll expressions
\begin{align}
a(\tau) \approx a_0 e^{-i\sqrt{\frac{V}{3}}\tau + \frac{V_{,\phi}^2}{12V}\tau^2}\,, \\
\phi(\tau) \approx \phi_{SP} + i \frac{V_{,\phi}}{\sqrt{3V}}\tau \label{slowrollphi}\,,
\end{align}
a solution that remains valid to our desired level of accuracy over ${\cal O}(\frac{1}{\sqrt{\epsilon}})$ Hubble times. The important observation is that, if we move in the Lorentzian time direction (imaginary $\tau$), then the imaginary part of $\phi$ does not vary any longer. Thus, if we manage to fix $\phi$ to take real values, it will remain real as the universe expands. The scale factor will remain (approximately) real if we match the late time solution to the early solution precisely at the equator of the $4-$sphere, which from the Lorentzian point of view will be the waist of the $4$-dimensional (quasi-)de Sitter hyperboloid -- this is achieved by taking 
\begin{align}
a_0 = i \sqrt{\frac{3}{V}}\,, \qquad \tau = \tau^R_{max} + i \tau^I = \sqrt{\frac{3}{V}}\frac{\pi}{2} + i \tau^I\,.
\end{align}
The scalar field thus remains real valued if we choose the imaginary part at the South Pole to be given by (cf. \eqref{slowrollphi})
\begin{align}
\phi_{SP}^I = - \frac{V_{,\phi}}{\sqrt{3V}}\tau^R_{max} = - \frac{V_{,\phi}}{V} \frac{\pi}{2} = - \sqrt{2\epsilon}\frac{\pi}{2}\,. \label{SPestimate}
\end{align}
Thus we see that a complex value of the scalar field is typically required at the South Pole, precisely in order to reach real field values at late times. The only exception is provided by an extremum of the potential, where $V_{,\phi}=0.$ Note that it is the imaginary part, rather than the argument, of the scalar that is constrained. This is because the scalar field appears as an exponent of the higher-dimensional metric \eqref{KKansatz}. The imaginary part of $\phi_{SP}$ is required to be small, of order ${\cal O}(\sqrt{\epsilon}),$ which justifies an assumption made earlier. 

We may also check the approximate analytical expressions given above by comparing with the numerical example shown in Figs. \ref{fig:excomplextime} and \ref{fig:exfields}. The potential is taken to be
\begin{align}
    V(\phi)=\frac{1}{100} - e^{-\phi}
\end{align}
Then, at $\phi=8,$ we expect from \eqref{SPestimate} a South Pole value $\phi_{SP}^I=-\frac{V_{,\phi}}{V}\frac{\pi}{2} \approx -0.054,$ while numerically we found $\phi_{SP}^I \approx -0.050.$ This confirms that our approximations are trustworthy at the level of a few percent, as expected.

We can now also see how this imaginary part compares with the bound on allowable metrics. First we will consider the case where the scalar arose as the volume modulus of a compactification. As we saw above, near the South Pole the metric is Euclidean to a good approximation, and only the scalar $\phi$ is complex valued. Thus, the $d=4$ version of the bound \eqref{Euclibound} applies in this case  
\begin{align}
|\textrm{Im}\,(\phi_{SP})| < \sqrt{\frac{D-4}{D-2}}\frac{\pi}{6\sqrt{2}} \quad \rightarrow \quad \frac{|V_{,\phi}|}{V} < \sqrt{\frac{D-4}{D-2}}\frac{1}{3\sqrt{2}}\,.
\end{align}
Up to a factor smaller than $2,$ this condition translates into $\frac{|V_{,\phi}|}{V} \lessapprox \frac{1}{3}.$ This means that only for $4$-dimensional potentials that are sufficiently flat do no-boundary solutions exist which, when seen from the higher-dimensional point of view, correspond to metrics on which quantum field theories may be consistently defined.

\begin{figure}[h]
	\centering
	\includegraphics[width=0.4\textwidth]{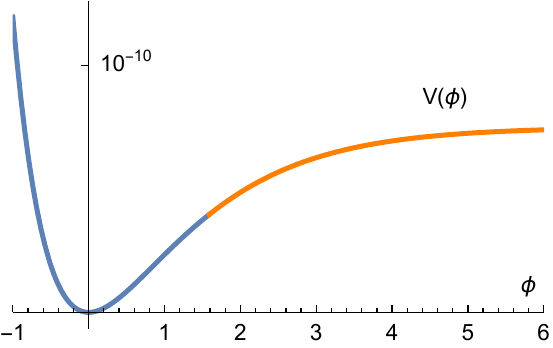}
	\caption{The effective scalar potential in the Starobinsky model of inflation. The orange part of the potential is the region in which the potential is flat enough to allow for no-boundary solutions with an allowable complex scalar field at the South Pole.}
	\label{fig:staro}
\end{figure}

When the scalar arises from a rescaling of the metric, the restriction on the potential is similar. The best known model of this kind is the Starobinsky model \cite{Starobinsky:1980te}, with action
\begin{align}
S = \frac{1}{2}\int d^4 x \sqrt{g} \left( R + \frac{1}{6M^2}R^2 \right)\,,
\end{align}
where $M$ is a mass scale, which needs to be set to about $10^{13}$ GeV in order to obtain a model compatible with CMB observations. The conformal transformation \eqref{phiR} corresponds to the identification $\sqrt{\frac{2}{3}}\phi = \ln (1+R/(3M^2)),$ and leads to the effective potential
\begin{align}
V(\phi) = \frac{3M^2}{4}\left( 1- e^{-\sqrt{\frac{2}{3}}\phi}\right)^2\,.
\end{align}
Near the South Pole of no-boundary solutions, \eqref{rsb} implies the bound
\begin{align}
|\textrm{Im}\,(\phi_{SP})| < \frac{\sqrt{6}}{8}\pi \quad \rightarrow \quad \frac{|V_{,\phi}|}{V} < \frac{\sqrt{6}}{4} \approx 0.61\,.
\end{align}
This means that the scalar potential must be flat enough in order for no-boundary solutions with an allowable complex scalar to exist. The bound above translates into the requirement that $\phi > 1.59,$ and we illustrate the corresponding range in Fig. \ref{fig:staro}. It is clear that this bound is not a strong restriction on the model -- the bound approximately corresponds to the requirement that inflation should last more than a couple of e-folds.

\begin{figure}[h]
	\centering
	\includegraphics[width=0.4\textwidth]{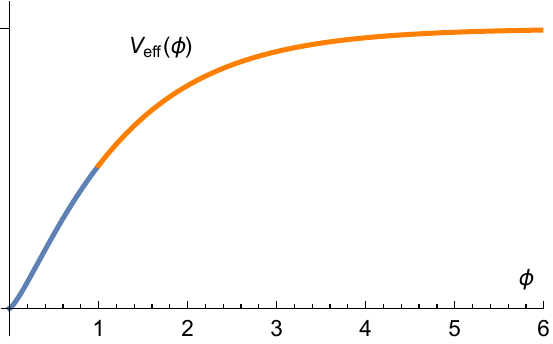}
	\caption{The effective scalar potential in the $8-$dimensional model \eqref{8d}. This model features both a rescaling of the metric due to a higher curvature term, and a flux compactification down to $4$ dimensions. Once again, the orange part of the potential is the region in which the potential is flat enough to allow for no-boundary solutions with an allowable complex scalar field at the South Pole.}
	\label{fig:staro2}
\end{figure}

Another example is the $8-$dimensional model studied in \cite{Lehners:2022mbd} (based on \cite{Ketov:2017aau,Otero:2017thw}), which starts with an action of the form
\begin{align}
 S &= \frac{1}{2}\int d^{8} x \sqrt{-\hat{g}}\left(\hat{R}+ \alpha \hat{R}^4 - \frac{1}{2\cdot4!}g_{YM}^2 F_{(4)}^2\right) \,.
\end{align}
The action thus contains both a higher curvature correction term and a flux term. Upon rescaling the metric to Einstein frame according to \eqref{metricconformal} and \eqref{phiR}, one obtains a scalar field with a potential that is asymptotically flat. This is a higher-dimensional analogue of the Starobinsky model. The theory is then compactified on a $4-$sphere, and this introduces a second scalar field. That second field is however stabilised by contributions from the flux term. For an appropriate range of fluxes, the overall potential is inflationary for the first scalar field $\phi,$ with an effective potential
\begin{align}
    V_{eff}(\phi) \approx V_{plateau} \left(1 - e^{-\sqrt{\frac{6}{7}}\phi} \right)^{\frac{4}{3}}\,. \label{8d}
\end{align} 
In fact, this is the attraction of the model: only for a certain range of fluxes do no-boundary solutions exist, so that the requirement that such universes come into existence automatically provides a selection principle on fluxes. In $8$ dimensions, the rescaling bound \eqref{rsb} reads
\begin{align}
|\textrm{Im}\,(\phi_{SP})| < \frac{\sqrt{42}}{16}\pi \quad \rightarrow \quad \frac{|V_{eff,\phi}|}{V_{eff}} < \frac{\sqrt{42}}{8} \approx 0.8\,.
\end{align}
This is not a particularly strong constraint on the potential, as it has to be inflationary in any case in order for no-boundary solutions to exist \cite{Hartle:2008ng,Lehners:2015sia}. We illustrate the allowed range graphically in Fig. \ref{fig:staro2}. Explicit examples of no-boundary solutions in this potetnial were discussed in detail in \cite{Lehners:2022mbd}. 

From all the examples above we can see that no-boundary solutions in sufficiently flat scalar potentials provide examples of solutions containing allowable complex scalar fields.


\section{Discussion} \label{sec:discussion}

The main observation of this work is that scalar fields that arise from the metric, be it via compactifications or rescalings, can take on complex values -- with the imaginary part being bounded -- without jeopardising the convergence properties of the original theory. This implies that the allowability criterion of Louko-Sorkin-Kontsevich-Segal-Witten may be extended somewhat, so as to allow for complex matter fields under the described circumstances.

It is important to note that this generalised setting still eliminates known pathological examples. For instance, the zero action wormholes considered by Witten \cite{Witten:2021nzp} are represented by a scale factor whose argument runs from $0$ to $\pi$. Thus, even if one adds a complex scalar field (with small backreaction), somewhere in this spacetime the allowability bound will be drastically violated, although the specific location where this occurs may change compared to the case where only real scalars are allowed. The same argument applies to proposed quantum bounces, such as those in \cite{Bramberger:2017cgf}: even though they contain a complex scalar field, the metric still runs through all possible values of its argument. Hence the imaginary part of the scalar cannot compensate for the complexity in the scale factor, and $\Sigma$ will exceed $\pi$ somewhere along these interpolating solutions.

An important class of solutions that our extended allowability criterion admits are no-boundary solutions with a scalar field. As we have shown, away from an extremum in the potential, these solutions require the scalar field to take on complex values. When the scalar has its origin in the metric, be it via compactification or conformal transformation, then the theory retains a convergent path integral as long as the scalar field does not become too complex. The precise bound depends on the context, but typically bounds the imaginary part by a factor of order unity (or slightly less). Given that no-boundary solutions require the imaginary part of the scalar to be of order ${\cal O}(\sqrt{\epsilon}),$ where $\epsilon$ is the slow-roll parameter, this poses no real conflict with the existence of no-boundary solutions, certainly not with realistic solutions. This shows that the no-boundary framework is consistent with a higher-dimensional context -- see also \cite{Jonas:2020pos,Lehners:2022mbd} for related works. 

At the moment, the relevance of the results presented here primarily concerns early universe cosmology. The no-boundary proposal remains the best understood and most consistent theory of initial conditions for the universe \cite{Jonas:2021xkx}. However, inflationary no-boundary solutions require a sufficiently flat potential in order to exist -- and as we saw, this fits well with an extended allowability criterion for complex scalar fields. However, this leads to some tension with supergravity and string theories, codified in the de Sitter swampland conjecture \cite{Obied:2018sgi}\footnote{Ekpyrotic no-boundary instantons \cite{Battarra:2014xoa} may well face an analogous tension with the swampland, see \cite{Lehners:2018vgi}.}. This conjecture essentially expresses the expectation that it will be difficult/impossible to find quasi-de Sitter solutions in classical supergravity or perturbative string theory. One way forward is then to consider non-perturbative corrections and loop effects, yielding e.g. higher curvature corrections. This is then precisely the framework we have analysed here, where complex scalars may find a natural home. Moreover, one may then envisage that allowability will help in selecting solutions.

It would be desirable to go beyond the examples studied here. We have only analysed examples of scalar fields arising from the metric in a rather simple way. There are many possible generalisations: one might look at more involved compactifications including shape moduli and vector fields, more general curvature corrections involving Ricci and Riemann tensors on top of the Ricci scalar, and start exploring non-cosmological solutions. These are all attractive avenues for future work.

\vspace{.5cm}

{\it Note added:} I have learned that Oliver Janssen, Joel Karlsson and Thomas Hertog independently arrived at similar conclusions regarding complex scalars in the no-boundary proposal \cite{private}.


\subsection*{Acknowledgments}

I would like to thank Oliver Janssen, Caroline Jonas, Thomas Hertog, Kelly Stelle and J\'{e}r\^{o}me Quintin for stimulating discussions and correspondence.
I gratefully acknowledge the support of the European Research Council (ERC) in the form of the ERC Consolidator Grant CoG 772295 ``Qosmology''.

\bibliographystyle{utphys}
\bibliography{ComplexScalarsBib}

\end{document}